\pdfoutput=1
\documentclass[twocolumn,prl,aps,longbibliography,showpacs,
superscriptaddress,10pt]{revtex4-1}

\usepackage{amsfonts,amssymb,dsfont}
\usepackage[sumlimits,intlimits]{amsmath}
\usepackage{graphics}
\usepackage{graphicx}
\usepackage{color}
\usepackage{mathrsfs}
\usepackage{textcomp}
\usepackage{verbatim}
\usepackage{hyperref}    % links
\usepackage{bm}          % bold math letters
\usepackage{braket}

\def\cH{\hat{\cal H}}

\def\cL{{\cal\hat L}}

\def\bn{{\bf n}}

\def\hbsigma{\hat{\boldsymbol \sigma}}

\def\holOne{\mathds{1}}

\def\hA{\hat A}

\def\hB{\hat B}

\def\hC{\hat C}
\def\hD{\hat D}

\def\hrho{\hat\rho}
\def\hX{\hat X}

\def\hV{\hat V}

\def\hsigma{\hat\sigma}

\newcommand{\env}[1]{\left< #1 \right>_{\text{env}}}

\begin{document}

\title{Out-of-time-order correlators in finite open systems}

\author{S.V.~Syzranov}
\affiliation{Joint Quantum Institute, NIST/University of Maryland, College Park, MD 20742, USA}
\affiliation{School of Physics and Astronomy, Monash University, Victoria 3800, Australia}
%\affiliation{Condensed Matter Theory Centre, Physics Department, University of Maryland, College Park, MD 20742, USA}

\author{A.V.~Gorshkov}
\affiliation{Joint Quantum Institute, NIST/University of Maryland, College Park, MD 20742, USA}
\affiliation{Joint Center for Quantum Information and Computer Science, NIST/University of Maryland, College Park, MD 20742, USA }

\author{V.~Galitski}
\affiliation{Joint Quantum Institute, NIST/University of Maryland, College Park, MD 20742, USA}

%\author{B.~Skinner}
%\affiliation{Massachusetts Institute of Technology, Cambridge, MA 02139 USA}

\date{\today}

\begin{abstract}
We study out-of-time order correlators (OTOCs) 
of the form $\langle\hA(t)\hB(0)\hC(t)\hD(0)\rangle$
for a quantum system weakly coupled to a dissipative environment. Such an open system
may serve as a model of, e.g., a small region
in a disordered interacting medium coupled to the rest of this medium considered as an environment. 
We demonstrate that for a system with discrete energy levels 
the OTOC saturates exponentially $\propto \sum a_i e^{-t/\tau_i}+const$
to a constant value at $t\rightarrow\infty$, in contrast with quantum-chaotic
systems which exhibit exponential growth of OTOCs.
Focussing on the case of a two-level system, we
calculate microscopically the  
decay times $\tau_i$ and the value of the saturation constant.
Because some OTOCs are immune to dephasing processes and some are not, such correlators
may decay on two sets of parametrically different
time scales related to inelastic transitions between the system levels and to pure dephasing processes, respectively.
In the case of a classical environment, the evolution of the OTOC can be mapped onto the evolution
of the density matrix of two systems coupled to the same dissipative environment.
%, which opens a new
%way to measure OTOCs, for example, in ensembles of spins or qubits in 
%dissipative environments. 
\end{abstract}

\maketitle

Quantum information spreading in a quantum system 
is often described by out-of-time-order
correlators (OTOCs) of the form 
\begin{align}
	K(t)=\left<{\hA(t)\hB(0)\hC(t)\hD(0)}\right>,
	\label{Kdef}
\end{align}
where $\hA$, $\hB$, $\hC$ and $\hD$ are %local 
Hermitian operators,
and $\langle\ldots\rangle$ is the average with respect to the initial state of the system.
Correlators of such form have been first 
introduced
by A.~Larkin and Y.N.~Ovchinnikov\cite{LarkinOvchnnikov}
in the context of disordered conductors,
where the correlator $\langle[p_z(t),p_z(0)]^2\rangle$ of particle momenta $p_z$
has been demonstrated to grow exponentially
$\propto e^{2\lambda t}$ for sufficiently long times $t$. 
The Lyapunov exponent $\lambda$ 
characterises the rate of divergence of two classical electron trajectories with slightly different
initial conditions and serves as a measure of quantum chaotic behaviour in a system. 

The concept of OTOC has revived\cite{Kitaev:talk} recently in the context of quantum information scrambling
and black holes,
motivating further studies of such quantities
(see, e.g., Refs.~\onlinecite{Maldacena:bound,SwingleChowdhury:LocScrambling,AleinerFaoroIoffe,Rosenbaum:rotor,PatelSachdev:FermiSurf}).
Despite not being measurable 
observables\footnote{
Because an OTOC involves
evolution backwards in time, measuring it 
requires either using a second copy of the system\cite{Yao:2copies,Knap:2copies,YungerHalpern:kindaReview}
or effectively reverting the sign of the Hamiltonian\cite{Swingle:measurement,ZhuHafeziGrover:measureClock,GarttnerRey:ionOTOC,Danshita:SYKmeasurement,Tsuji:measurement,Li:NMRmeas},
possible, e.g., in spin systems using spin-echo-type techniques or ancilla qubits},
OTOCs (\ref{Kdef}) characterise the spreading of quantum information 
%(quantum information scrambling) 
and the sensitivity of the system
to the change of the initial conditions.
It is also expected that
OTOCs may be used\cite{Fan:OTOCMBL,Chen:MBLscrambling,SwingleChowdhury:LocScrambling,HuangChen:OTOCMBL,Sachdev:ExpWeakDisorder}
to distinguish between
many-body-localised and many-body-delocalised states\cite{BAA}
of disordered interacting systems.

So far the studies of quantum chaos and information scrambling have been focussing on closed quantum systems.
In reality, however, each system is coupled to a noisy environment, which leads to decoherence and affects
information spreading. Moreover, a
sufficiently strongly disordered interacting system
may be separated into a small subsystem, of the size of the single-particle localisation length or
a region of quasi-localised states,
coupled to the rest of the system considered as environment.
In this paper we analyse out-of-time order correlators in a quantum system weakly coupled to a dissipative environment.

{\it Phenomenological picture in a strongly disordered material.}
A system with localised single-particle states and weak short-range interactions
exhibits insulating behaviour at low temperatures\cite{BAA}.
Local physical observables in such  a system
are strongly correlated only on
short length scales, and their properties may be understood 
by considering a single ``localisaton cell'', particle states in a region
of space of the size of order of the localisation length $\xi$, which may be considered weakly coupled
to the rest of the system.

\begin{figure}[b]
	\centering
	\includegraphics[width=0.99\columnwidth]{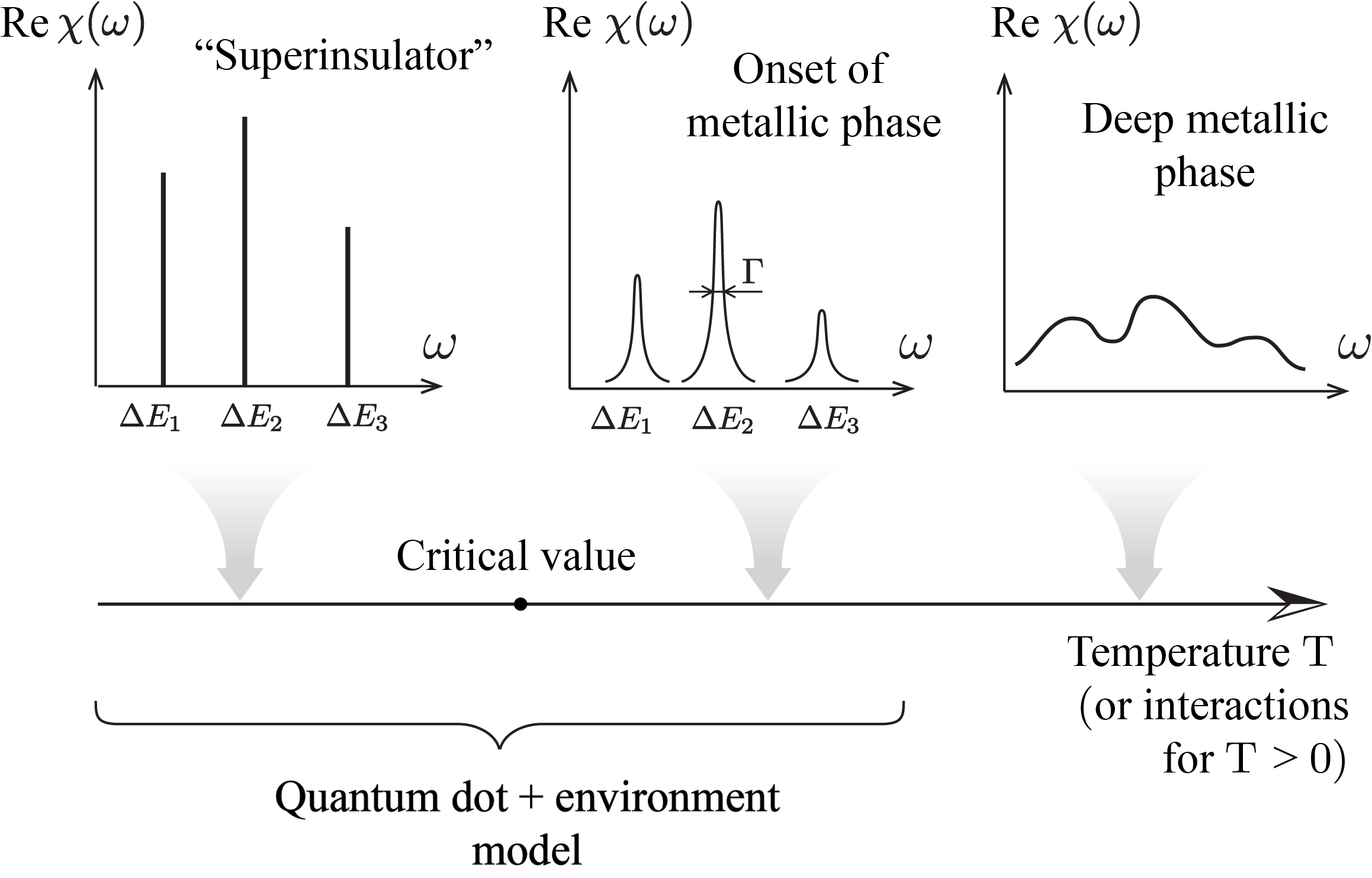}
	\caption{\label{GammaRegimes}
	Response of a localisation cell in a strongly disordered interacting system
	for various temperatures $T$ (or interaction strengths for $T>0$).
	$\Delta E_i$ are the energy gaps between the many-body levels in the cell, and
	$\Gamma$ is the level width.	
	}
\end{figure}

The energy spectrum of the localisation cell may be probed via response functions 
of local operators in the cell, e.g., the response function
$\chi(\omega)=\sum_{\alpha,\beta}\frac{(f_\alpha-f_\beta)|Q_{\alpha\beta}|^2}{E_\alpha-E_\beta+\omega+i0}$
of the charge
$Q$ in a region inside the cell to the voltage in this region, where $E_\alpha$ and $E_\beta$ are
the energies of many-body states and $f_\alpha$ is their distribution function.
For temperatures smaller than a critical value, 
quasiparticles in the system have zero decay 
rate\footnote{When studying response functions, a system in the insulating state should be assumed coupled
to an external bath, with the value of the coupling sent to zero at the end of the calculation\cite{BAA}. The quasiparticle
decay rate is then given by the bath strength.} 
(``superinsulating'' regime\cite{BAA}), and the
system thus responds only at a discrete set of frequencies
$\omega=E_i-E_j$, determined by the energy gaps between many-body states,
as shown in Fig.~\ref{GammaRegimes}. The OTOC (\ref{Kdef})
in this regime oscillates $K(t)\propto \sum_n a_n e^{i\omega_n t}$ with a discrete set
of frequencies $\omega_n=E_{i_n}+E_{i_n^\prime}-E_{j_n}-E_{j_n^\prime}$.

When the temperature (or the interaction strength at a given temperature) exceeds a critical value,
the levels and response functions get broadened (``metallic'' phase\cite{BAA}),
as illustrated in Fig.~\ref{GammaRegimes}, becoming smoother with increasing
temperature and/or interactions. Near the superinsulator-metal transition the characteristic level
width $\Gamma$ is significantly smaller than the gaps between levels, and the localisation cell
may be considered as an open system weakly coupled to a dissipative environment.
The same model may be applied also to a strongly disordered material
with an external bath, such as a system of phonons, which provide a finite level width $\Gamma$
at all finite temperatures.
The local operators $\hA$, $\hB$, $\hC$ and $\hD$ in Eq.~(\ref{Kdef}) do not
necessarily act on states in one localisation cell, but may involve states in several cells close to each other.
These cells may still be considered as a single quantum dot in a noisy environment
so long as the level spacing in the dot exceeds the level width.
Such a model of an open quantum dot
may be also realised directly, e.g., using superconducting qubits or trapped cold atoms.

\begin{figure}[b]
	\centering
	\includegraphics[width=0.27\textwidth]{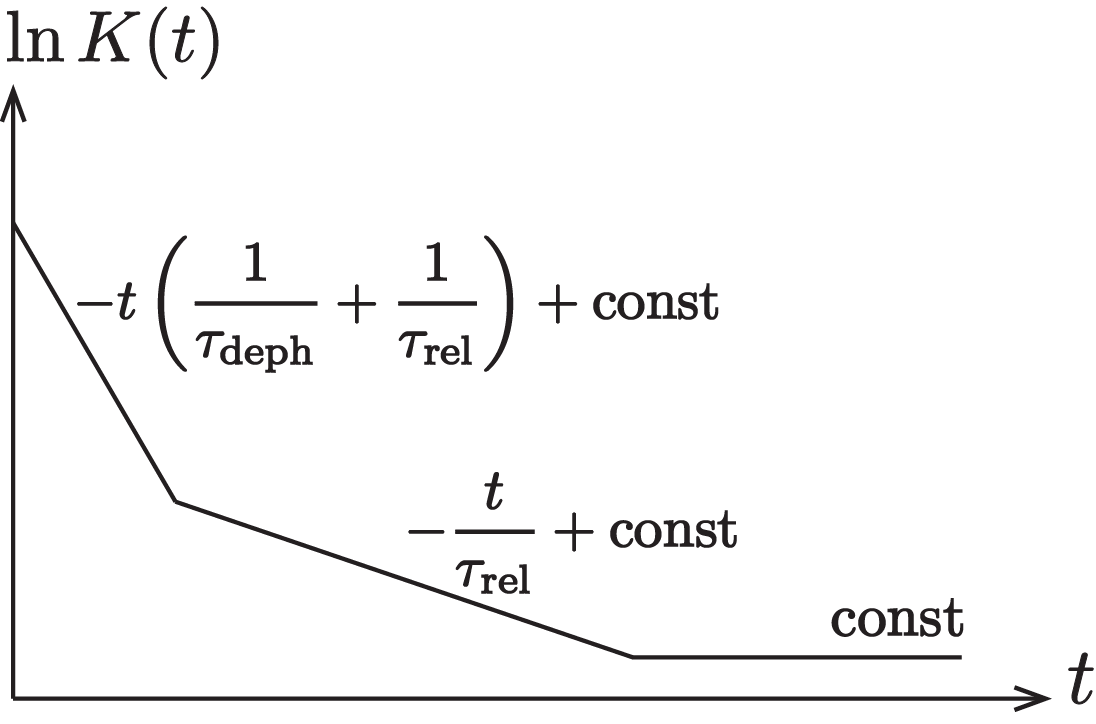}
	\caption{
	Time dependence of the out-of-time-order
	correlator (\ref{Kdef}) in a system with significantly larger dephasing rate
	than relaxation. $\tau_{\text{deph}}$ and $\tau_{\text{rel}}$ are the (longest) characteristic
	times of dephasing and inelastic relaxation.}
	\label{Fig:RelPlot}
\end{figure}

{\it Summary of the results.}
We demonstrate that, for a system with discrete non-degenerate levels $E_n$, correlator
(\ref{Kdef}) at long times $t$ exponentially saturates to a constant value,
$K(t)\propto \sum a_n e^{i\omega_n t}e^{-t/\tau_n}+const$, and calculate microscopically
the value of the constant and the relaxation times $\tau_n$ as a function of the environment
spectral function and the matrix elements of the system-environment coupling.
Depending on the choice of the operators $\hA$, $\hB$, $\hC$ and $\hD$, the
saturation value may be finite or zero.
OTOCs relax due to both inelastic transitions between the system's levels
and pure dephasing processes, which are caused by slow fluctuations of the energies $E_n$.
While some OTOCs are immune to dephasing processes, a generic correlator has components
both sensitive and insensitive to dephasing and thus decays on two sets of parametrically
different scales related to dephasing and relaxation respectively, as shown in Fig.~\ref{Fig:RelPlot}.

Our results indicate, in particular, that a disordered system of interacting particles 
cannot exhibit quantum chaotic behaviour 
if the typical single-particle level splitting $\delta_\xi$ 
in a volume of linear size $\xi$ (localisation length) exceeds
the dephasing rate and the rate of inelastic transitions due to interactions and/or phonons.
Correlators (\ref{Kdef}) in this system can only 
saturate to constant values at $t\rightarrow\infty$, in contrast with quantum-chaotic systems
which display exponential growth of OTOCs with time.
Our results thus suggest that chaotic behaviour in a disordered interacting system requires either
the presence of delocalised single-particle states or sufficiently strong interactions 
or, e.g., a phonon bath,
which would lead to the
quasiparticle decay rate exceeding the level spacing $\delta_\xi$.

For a classical environment, the evolution of an OTOC (\ref{Kdef}) in an open system 
may be mapped onto
the evolution of the density matrix of two systems coupled to the same environment,
which allows one to measure OTOCs by observing the
correlations between two systems in a noisy environment, such as spins in a random time-dependent
magnetic field.

{\it Model.} We consider a system with discrete {\it non-degenerate}
energy levels $E_n$ coupled to a dissipative environment and
described by the Hamiltonian
\begin{align}
	\cH=\cH_0 + \hat V \hat X + \cH_{\text{bath}}(\hat X),
	\label{Ham}
\end{align}
where $\cH_0=\sum_n E_n\ket{n}\bra{n}$ is the Hamiltonian of the system,
$\cH_{\text{bath}}(\hat X)$-- the Hamiltonian of the environment, and $\hat V \hat X$
is the coupling between the system and the environment, where the operator
$\hV=\sum_{n,m}V_{nm}\ket{n}\bra{m}$ acts on the system degrees of freedom, and $\hX$ is
an environment variable which commutes with the system degrees of freedom.

%{\it Microscopic equations of evolution.}
To compute the OTOC (\ref{Kdef}), where the operators
$\hA$, $\hB$, $\hC$ and $\hD$ act on the system variables,
it is convenient to decompose it
as $K=K_{m_1m_2,n_1n_2}A_{n_1m_1}C_{n_2m_2}$ (summation over repeated indices implied),
where $A_{n_1m_1}$ and $C_{n_2m_2}$,
are the matrix elements of the operators $\hA$ and $\hC$, and
\begin{align}
	K_{m_1m_2,n_1n_2}=\left<{\ket{n_1}\bra{m_1}(t)\hB(0)\ket{n_2}\bra{m_2}(t)\hD(0)}\right>,
	\label{KelemDef}
\end{align}
where $\langle\ldots\rangle$ is the averaging with respect to both the system and environment states. 

%As we demonstrate below, the evolution of the elements $K_{m_1m_2,n_1n_2}$ is similar to that
%of the density matrix elements $\rho_{m_1m_2,n_1n_2}$ of a compound system consisting of two copies
%of the original system coupled to the same environment (where $n_i$ and $m_i$ label the states
%of the $i$-th copy respectively).

In the limit of a vanishing system-environment coupling $\hV$, the correlators (\ref{KelemDef})
oscillate with time, $K_{m_1m_2,n_1n_2}\propto e^{i(E_{n_1}+E_{n_2}-E_{m_1}-E_{m_2})t}$.
A finite coupling between the system and the environment leads to dissipation and relaxation
processes and thus to the decay of the elements $K_{m_1m_2,n_1n_2}$.
For a weak coupling considered in this paper, the characteristic decay times
of the OTOCs significantly exceed the correlation time of the environment degrees of freedom, i.e.
of the function $S(t-t^\prime)=\langle\hX(t)\hX(t^\prime)\rangle_{\text{env}}$, and the evolution
of the elements is described by a system of Markovian Bloch-Redfield\cite{Slichter:book} equations
(see Supplemental Material for the microscopic derivation)
of the form
\begin{align}
	&\partial_t K_{m_1m_2,n_1n_2}
	=
	\nonumber\\
	& i(E_{n_1}+E_{n_2}-E_{m_1}-E_{m_2})K_{m_1m_2,n_1n_2}
	\nonumber\\
	& -\sum_{m_1^\prime, m_2^\prime,n_1^\prime, n_2^\prime} 
	\Gamma_{m_1m_2,n_1n_2}^{m_1^\prime m_2^\prime,n_1^\prime n_2^\prime}
	K_{m_1^\prime m_2^\prime,n_1^\prime n_2^\prime}.
	\label{KelemEvol}
\end{align}

From the definition of the elements (\ref{KelemDef}) it follows that
\begin{align}
	\sum_{m,n} K_{nm,nm}=\left<{\hB(0)\hD(0)}\right>
	=\text{const}.
	\label{DiagConserv}
\end{align}
%The conservation law (\ref{DiagConserv}) is similar to the conservation of the sum of the diagonal
%elements of the density matrix of a system consisting of two subsystems with states labelled by
%$n$ and $m$. 
Eq.~(\ref{DiagConserv}) may be also derived from the microscopic equations of evolution,
as shown in Supplemental Material.

Due to the smallness of the decay rates $\Gamma_{m_1m_2,n_1n_2}^{m_1^\prime m_2^\prime,n_1^\prime n_2^\prime}$
in Eq.~(\ref{KelemEvol}),
the evolution of each element $K_{m_1 m_2,n_1 n_2}$ is affected only by the elements $K_{m_1^\prime m_2^\prime,n_1^\prime n_2^\prime}$
with the same oscillation frequency $E_{n_1}+E_{n_2}-E_{m_1}-E_{m_2}$ (secular approximation).
In this paper we consider systems with sufficiently non-degenerate energy spectra;
if two elements oscillate with the same frequency, they may be different only by permutations
of indices $n_1$ and $n_2$ and/or $m_1$ and $m_2$.

For a generic $N$-level system there are $2N^2-N$ elements (\ref{KelemDef})
with zero energy gaps $E_{n_1}+E_{n_2}-E_{m_1}-E_{m_2}$ (with $m_1=n_1$, $m_2=n_2$ and/or $m_1=n_2$, $m_2=n_1$).
These elements are immune to dephasing, i.e. to the accumulation of
random phases caused by slow fluctuations
of the energies $E_{n_i}$.
Such vanishing of dephasing is similar to that in decoherence-free
subspaces\cite{Zanardi:decoherencefree,WuLidar:decoherencefree} of multiple-qubit systems.
We emphasise, however, that even dephasing-immune
correlators in general decay at long times due to the
environment-induced inelastic transitions between the levels (relaxation processes).

A generic OTOC (\ref{Kdef})  
includes components
both sensitive and insensitive to dephasing, as well as a component independent of time, which exists due to 
the conservation law (\ref{DiagConserv}). 
For an environment with a smooth spectral function on the scale of the characteristic level splitting,
the characteristic decay rate of the dephasing-immune components may be estimated as $1/\tau_{\text{rel}}
\sim V_\bot^2S(\Delta E)$, where $V_\bot$ is the typical off-diagonal matrix element of the perturbation $\hV$
and $\Delta E$ is the characteristic level spacing.
The other components decay with the characteristic rate $1/\tau_{\text{deph}}+1/\tau_{\text{rel}}$,
where $1/\tau_{\text{deph}}\sim V_\parallel^2S(0)$ is the characteristic dephasing rate,
where $V_\parallel$ is the typical diagonal matrix element of the perturbation $\hV$.
As a result, the decay of the OTOC consist of three stages,
corresponding to these characteristic times, as illustrated in Fig.~\ref{Fig:RelPlot}.

{\it Two-level system.} In order to illustrate the meaning of these time scales and the related phenomena,
we focus below on the case of a two-level system, equivalent to a spin-$1/2$ in a random magnetic field
(for the microscopic analysis of OTOCs in the generic case of a multi-level system see Supplemental Material),
described by the Hamiltonian 
\begin{align}
	\cH=\frac{1}{2}B\hsigma_z +
	\frac{1}{2}
	\hbsigma
	\bn\hX
	+\cH_{bath}(\hX),
	\label{HamSpin}
\end{align}
where $\hbsigma$ is a vector of Pauli matrices and $\bn$ is a constant unit vector, the direction
of the fluctuations of the magnetic field.

The dissipative environment induces transitions $\ket{\uparrow}\rightarrow\ket{\downarrow}$ with the rate
$\Gamma_\downarrow=\frac{1}{4}(n_x^2+n_y^2)S(B)$, as well as the opposite transitions 
$\ket{\downarrow}\rightarrow\ket{\uparrow}$ with the rate $\Gamma_\uparrow=\frac{1}{4}(n_x^2+n_y^2)S(-B)$,
where $S(\omega)$ is the environment spectrum, the Fourier-transform
of $S(t-t^\prime)=\langle \hX(t)\hX(t^\prime)\rangle_{\text{env}}$. Weak fluctuations of the magnetic field
in the longitudinal direction lead to dephasing with the rate $\Gamma^\phi=\frac{1}{2}n_z^2S(0)$. 
We focus below on the long-time dynamics of the system and assume for simplicity
that the rate $\Gamma^\phi$ of pure dephasing significantly
exceeds the rates $\Gamma_\uparrow$ and $\Gamma_\downarrow$ of inelastic
transitions between the levels of the spin; in the opposite case, all OTOC decay  rates
are of the same order of magnitude.

The OTOCs $K_{\uparrow\uparrow,\downarrow\downarrow}$ and $K_{\downarrow\downarrow,\uparrow\uparrow}$
oscillate with frequencies $\pm 2(E_\downarrow-E_\uparrow)=\mp 2B$ and have dephasing rate
$4\Gamma^\phi$, the same as $\pm1$-projection states of a spin-$1$ in magnetic field $B$,
\begin{align}
	K_{\uparrow\uparrow,\downarrow\downarrow},K_{\downarrow\downarrow,\uparrow\uparrow}\propto 
	e^{\mp 2i Bt}e^{-4\Gamma^\phi t},
\end{align}
where we have neglected the small relaxation rates $\Gamma_{\uparrow,\downarrow}\ll\Gamma^\phi$.

There are 8 elements (\ref{KelemDef}) which
correspond to 3 spin indices pointing in one direction and one spin index pointing in the opposite direction.
These elements oscillate with frequencies $\pm B$ and have the same dephasing rate as a spin-$1/2$,
\begin{align}
	K_{\uparrow\downarrow,\downarrow\downarrow},K_{\downarrow\uparrow,\uparrow\uparrow},
	K_{\downarrow\downarrow,\uparrow\downarrow},\ldots\propto e^{-\Gamma^\phi t}.
\end{align}

\begin{figure}[htbp]
	\centering
	\includegraphics[width=0.8\columnwidth]{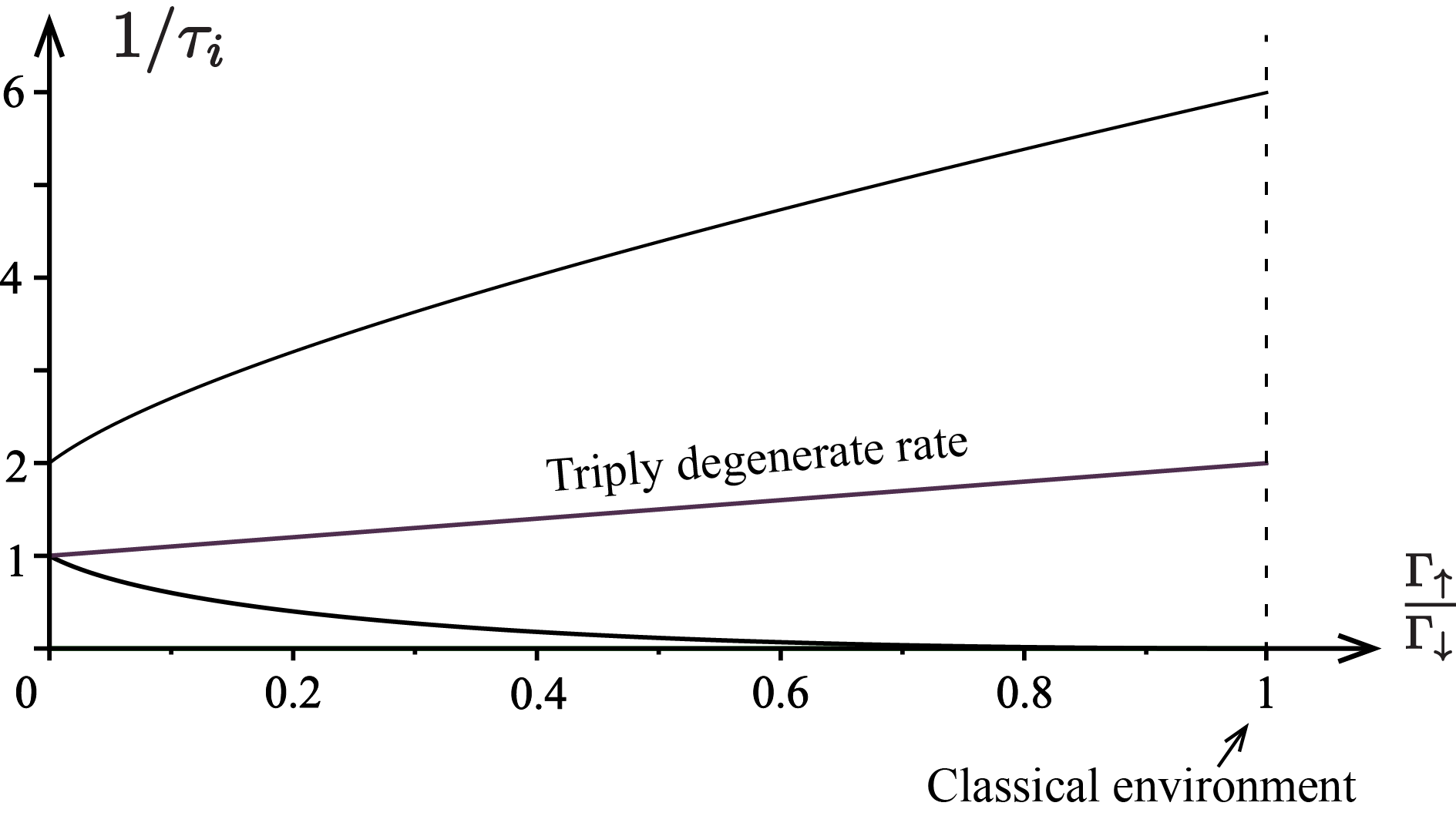}
	\caption{Non-zero rates (in units $\Gamma_\downarrow$) of long-time decay
	of out-of-time-order correlators
	in a two-level system as a function of the ratio $\Gamma_\uparrow/\Gamma_\downarrow$
	of the transition rates between the system levels.
	\label{RatesPlot}}
\end{figure}

The behaviour of OTOCs at long times $t\gg 1/\Gamma^\phi$ is determined by the components with a vanishing
frequency $E_{n_1}+E_{n_2}-E_{m_1}-E_{m_2}$ of coherent oscillations, because such components
are insensitive to dephasing. For a spin-1/2, their evolution is described by the
system of equations (as follows from the generic master equations for a multi-level system derived in
Supplemental Material)
\begin{widetext}
\begin{align}
	\partial_t
	\left(
	\begin{array}{c}
		K_{\downarrow\uparrow,\uparrow\downarrow}\\
		K_{\uparrow\downarrow,\downarrow\uparrow}\\
		K_{\uparrow\downarrow,\uparrow\downarrow}\\
		K_{\downarrow\uparrow,\downarrow\uparrow}\\
		K_{\uparrow\uparrow,\uparrow\uparrow} \\
		K_{\downarrow\downarrow,\downarrow\downarrow}
	\end{array}
	\right)
	=
	\left(
	\begin{array}{cccccc}
		-\Gamma_\downarrow-\Gamma_\uparrow & 0 & -\Gamma_\downarrow & -\Gamma_\downarrow & \Gamma_\downarrow & \Gamma_\downarrow \\
		0 & -\Gamma_\downarrow-\Gamma_\uparrow & -\Gamma_\uparrow & -\Gamma_\uparrow & \Gamma_\uparrow & \Gamma_\uparrow \\
		-\Gamma_\uparrow & -\Gamma_\downarrow & -\Gamma_\downarrow-\Gamma_\uparrow & 0 & \Gamma_\downarrow & \Gamma_\uparrow \\
		-\Gamma_\uparrow & -\Gamma_\downarrow & 0 & -\Gamma_\downarrow-\Gamma_\uparrow & \Gamma_\downarrow & \Gamma_\uparrow \\
		\Gamma_\uparrow & \Gamma_\downarrow & \Gamma_\uparrow & \Gamma_\uparrow & -2\Gamma_\downarrow & 0 \\
		\Gamma_\uparrow & \Gamma_\downarrow & \Gamma_\downarrow & \Gamma_\downarrow & 0 & -2\Gamma_\uparrow 
	\end{array}
	\right)
	\left(
	\begin{array}{c}
		K_{\downarrow\uparrow,\uparrow\downarrow}\\
		K_{\uparrow\downarrow,\downarrow\uparrow}\\
		K_{\uparrow\downarrow,\uparrow\downarrow}\\
		K_{\downarrow\uparrow,\downarrow\uparrow}\\
		K_{\uparrow\uparrow,\uparrow\uparrow} \\
		K_{\downarrow\downarrow,\downarrow\downarrow}
	\end{array}
	\right).
	\label{SpinEvolSystem}
\end{align}
\end{widetext}
The rates of the long-time decay of OTOCs are given by the eigenvalues of the matrix in
Eq.~(\ref{SpinEvolSystem}) (with minus sign) and are shown (except for the zero eigenvalue) in Fig.~\ref{RatesPlot}.
Such a matrix always has a zero eigenvalue, due to the conservation law (\ref{DiagConserv}).
The system also has a triply degenerate decay rate $\Gamma_\uparrow+\Gamma_\downarrow$.
The other two decay rates are given by $\frac{1}{2}
\left[3\Gamma_\uparrow+3\Gamma_\downarrow\pm\left(\Gamma_\uparrow^2+34\Gamma_\uparrow\Gamma_\downarrow
+\Gamma_\downarrow^2\right)^\frac{1}{2}\right]$.

At long times $t\rightarrow\infty$ the correlator (\ref{Kdef}) saturates to a constant value
determined by the projection of the OTOC (\ref{Kdef}) on the zero-decay-rate mode,
\begin{align}
	K(t\rightarrow\infty)
	=
	\frac{1}{\sqrt{2\Gamma_\uparrow^2+2\Gamma_\downarrow^2}}
	&\left(
	\Gamma_\downarrow A_{\uparrow\downarrow}C_{\downarrow\uparrow}
	+\Gamma_\uparrow A_{\downarrow\uparrow}C_{\uparrow\downarrow}
	\right.
	\nonumber\\
	&
	\left.
	+\Gamma_\uparrow A_{\uparrow\uparrow}C_{\uparrow\uparrow}
	+\Gamma_\downarrow A_{\downarrow\downarrow}C_{\downarrow\downarrow}
	\right).
	\label{SaturationValue}
\end{align}
While we assumed a small inelastic relaxation rate in comparison with the dephasing rate,
we emphasise that the result (\ref{SaturationValue}) for the saturation value
of the OTOC holds for an arbitrary ratio of dephasing and relaxation rates.

%%%%%%%%%%%%%%%%%%%%%%%%%%%%%%%%%%%%%%%%%%%%%%%%%%%%%%%%%%%%%%%%%%%%%%%%%%%%%%%%%%%%%%%%%%%%%%%%%%%%%%%%
%%%%%%%%%%%%%%%%%MAPPING%%%%%%%%%%%%%%%%%%%%%%%%%%%%%%%%%%%%%%%%%%%%%%%%%%%%%%%%%%%%%%%%%%%%%%%%%%%%%%%%

{\it Mapping to the evolution of two systems for a classical environment.}
The evolution of the OTOCs (\ref{KelemDef}) is similar to that of the density-matrix elements
\begin{align}
	\rho_{m_1m_2,n_1n_2}
	=\left<\left<\ket{m_1}\bra{n_1}(t)\right>_{\text{Sys}_1}\left<\ket{m_2}\bra{n_2}(t)\right>_{\text{Sys}_2}\right>_X
	\label{rhoTwoSyst}
\end{align}
of a compound system consisting of two identical
subsystems (``Sys$_1$'' and ``Sys$_2$'') coupled to the same dissipative
environment, where $m_1$ and $n_1$ and $m_2$ and $n_2$ in Eq.~(\ref{rhoTwoSyst}) are the
states of the first and the second subsystems respectively, 
$\ket{m_i}\bra{n_i}(t)$ is an operator in the interaction representation,
and $\langle\ldots\rangle_X$ is the averaging
with respect to the environment degrees of freedom.
The Hamiltonian of such a compound system is given by
\begin{align}
	\cH=\cH_0\otimes\holOne + \holOne\otimes\cH_0
	+\left(\hV\otimes\holOne + \holOne\otimes\hV\right)\hX+\cH_{\text{bath}}(\hat X),
	\label{HamTwoSyst}
\end{align}
where $\ldots\otimes\ldots$ is the product of the subsystem subspaces; $\cH_0$ and $\hV$ are the Hamiltonian
of each subsystem and its coupling to the environment, and the environment variable $\hX$ commutes with all degrees
of freedom of subsystems ``Sys$_1$'' and ``Sys$_2$''.

The evolution of the elements (\ref{KelemDef}) and (\ref{rhoTwoSyst}) is described
by similar Markovian master equations (see Supplemental Material for microscopic derivation).
In particular, in the limit of a classical environment ($\env{\hX(t)\hX(t^\prime)}=\env{\hX(t^\prime)\hX(t)}$),
the evolution of OTOCs (\ref{KelemDef}) can be mapped exactly onto that of
the density matrix (\ref{rhoTwoSyst}) of two systems coupled
to this environment, as follows from the 
definitions of these quantities. The conservation law~(\ref{DiagConserv}) is mapped then  onto the
conservation of the trace of the density matrix of a compound system consisting of two subsystems.

In the limit of a classical environment, the spectral function is even, $S(\omega)=S(-\omega)$,
the relaxation rate $i\rightarrow j$ for each pair of levels $i$ and $j$ in a system
matches the reverse rate $j\rightarrow i$.
In particular, in the case of a two-level system $\Gamma_\uparrow=\Gamma_\downarrow=\Gamma$, and
the OTOC has three decay rates
at long times $t\gg{\Gamma^\phi}^{-1}$:
$\Gamma_1=6\Gamma$, $\Gamma_2=2\Gamma$ (triply degenerate) and $\Gamma_3=0$ (doubly degenerate),
as shown in Fig.~\ref{RatesPlot}. Due to the mapping, these rates match the decay 
rates of pair-wise correlators of observables
in, e.g.,
an ensemble of spins in a uniform random magnetic field and thus may be conveniently measured in such ensembles. 

We emphasise that the mapping between an OTOC and the evolution of two subsystems coupled to the same
classical environment holds
for an arbitrary system-environment coupling but not only in the limit of a weak coupling
considered in this paper. This mapping suggests a way for measuring OTOCs in generic systems
in the presence of classical environments through observing correlators 
$\left<\left<\hA(t)\right>_{\text{Sys}_1}\left<\hC(t)\right>_{\text{Sys}_2}\right>_X$
of observables $\hA$ and $\hC$ between two systems.

{\it Discussion.} We computed OTOCs in a system
weakly coupled to a dissipative environment and demonstrated that they saturate to a constant value
at long times.
Because such an open system may serve as a model of a small region in a disordered interacting medium
(in the presence or in the absence of a phonon bath),
this suggests the absence of a chaotic behaviour in strongly disordered materials.
While our result applies to weakly-conducting and insulating materials, for which the system-environment coupling
may be considered small, we leave it for a future study whether non-chaotic behaviour persists in systems
strongly coupled to the environment (corresponding to an effectively continuous energy spectrum of a localisation
cell). For a classical environment, the evolution of an OTOC matches
the evolution of correlators of observables between two identical systems coupled to the same environment,
which may be used for measuring OTOCs in open systems in classical environments.
The possibility to develop a similar measurement 
method for the case of a quantum environment is another question which
deserves further investigsation.

\begin{acknowledgments}
We have benefited from discussions with Yidan Wang.
V.G. and S.V.S. were supported by  US-ARO (contract No. W911NF1310172), NSF-DMR 1613029 and Simons Foundation;
A.V.G. and S.V.S. acknowledge support by NSF QIS, AFOSR, NSF PFC at JQI, ARO MURI, ARO and ARL CDQI.
%by AFOSR, NSF QIS, ARO MURI, ARO, ARL CDQI and NSF PFC at Joint Quantum Institute.
S.V.S. also acknowledges the hospitality of
School of Physics and Astronomy at Monash University, where a part of this work
was completed. 
\end{acknowledgments}

%%%%%%%%%%%%%%%%%%%%%%%%%%%%%%%%%%%%%%%%%%%%%%%%%%%%%%%%%%%%%%%%%%%%%%%%%%%%%%%%%%%%%%%%%%%%%%%%%%%%%%%%%%%%%%%%
%%%%%%%%%%%%%%%%BIBLIOGRAPHY AND SUPPLEMENTAL%%%%%%%%%%%%%%%%%%%%%%%%%%%%%%%%%%%%%%%%%%%%%%%%%%%%%%%%%%%%%%%%%%%


%%%%%%%%%%%%%%%%%%%%%%%%%%%%%%%%%%%%%%%%%%%%%%%%%%%%%%%%%%%%%%%%%%%%%%%%%%%%%%%%%%%%%%%%%%%%%%%%%%%%%%%%
%%%%%%%%%%%%%%%Supplemental Material%%%%%%%%%%%%%%%%%%%%%%%%%%%%%%%%%%%%%%%%%%%%%%%%%%%%%%%%%%%%%%%%%%%%

\newpage

\renewcommand{\theequation}{S\arabic{equation}}
\renewcommand{\thefigure}{S\arabic{figure}}
\renewcommand{\thetable}{S\arabic{table}}
\renewcommand{\thetable}{S\arabic{table}}
\renewcommand{\bibnumfmt}[1]{[S#1]}

\setcounter{equation}{0}
\setcounter{figure}{0}
\setcounter{enumiv}{0}

\onecolumngrid
\newpage
\begin{center}
\textbf{\large Supplemental Material for \\
``Out-of-time-order correlators in finite open systems''}
\end{center}
\vspace{2ex}
%\twocolumngrid

\subsection{Master equations for the density matrix in an open system} 

%Before deriving microscopically master equations for the evolution of out-of-time-order correlators in
%open systems, we summarise 

{\it Off-diagonal elements.}
For a system with non-degenerate energy levels weakly coupled to a dissipative environment,
with the Hamiltonian given by Eq.~(\ref{Ham}),
the off-diagonal entries $\rho_{mn}$ of the density matrix satisfy Bloch-Redfield master equations
(see, e.g., Ref.~\onlinecite{Slichter:book})
\begin{align}
	\partial_t \rho_{mn}= i(E_{mn}+i\Gamma_{mn})\rho_{mn},
	\label{RhoOffDiagMaster}
\end{align}
where $E_{mn}=E_m-E_n$ is the frequency of coherent oscillations for
an isolated system, and the complex quantity
\begin{align}
	\Gamma_{mn}=-i\int\frac{d\omega}{2\pi}\sum_k 
	\left(\frac{S(\omega)|V_{mk}|^2}{\omega-E_{mk}-i0}
	+\frac{S(-\omega)|V_{nk}|^2}{\omega-E_{kn}-i0}\right)
	+iV_{mm}V_{nn}\int\frac{d\omega}{2\pi}
	\frac{S(\omega)+S(-\omega)}{\omega-i0}
	\label{GammaMN}
\end{align}
accounts for the effects of the environment, where
$S(\omega)$ is the Fourier-transform
of the correlation function
$S(t-t^\prime)=\langle \hX(t)\hX(t^\prime)\rangle_{env}
=\int\frac{d\omega}{2\pi}e^{-i\omega(t-t^\prime)}S(\omega)$ of the environment degree of freedom $\hX(t)$.
%We note that in the above sums there are terms with $k=n$ and $k=m$ which, in addition to the last term,
%contribute to
%the dephasing rates.

The quantity $\Gamma_{mn}$, given by Eq.~(\ref{GammaMN}), may be decomposed as
\begin{align}
	\Gamma_{mn}=
	\frac{1}{2}\sum_{k\neq m}\Gamma_{m\rightarrow k}^{rel}
	+\frac{1}{2}\sum_{k\neq n}\Gamma_{n\rightarrow k}^{rel}
	-i\delta E_m + i\delta E_n
	+\Gamma_{mn}^{deph},
\end{align}
where
\begin{align}
	\Gamma_{n\rightarrow k}^{rel}=|V_{nk}|^2 S(E_{n}-E_{k})
	\label{GammaRel}
\end{align}
is the rate of environment-induced transitions (relaxation) from level $n$ to level $k$,
\begin{align}
	\Gamma_{mn}^{deph}=\frac{1}{2}(V_{nn}-V_{mm})^2S(0)
	\label{GammaDeph}
\end{align}
is the pure dephasing rate, and
\begin{align}
	\delta E_m=\sum_{k\neq m}|V_{mk}|^2\int\frac{d\omega}{2\pi}\frac{S(\omega)}{E_{m}-E_{k}-\omega}
	\label{LambShift}
\end{align}
is the shift of the energy of the $m$-th level due to the interaction with environment (Lamb shift). 
The relaxation rate between two levels $n$ and $k$, Eq.~(\ref{GammaRel}), is determined by the environment
spectrum $S(\omega)$ at frequency $\omega=E_{nk}$ equal to the energy gap between these levels, while
the dephasing rate (\ref{GammaDeph}) is determined by the low-frequency properties of the environment.

{\it Diagonal elements.}
The dynamics of the diagonal elements of the density matrix is described by the equations
\begin{align}
	\partial_t \rho_{nn}=-\rho_{nn}\sum_k\Gamma_{n\rightarrow k}^{rel} 
	+\sum_k\rho_{kk}\Gamma_{k\rightarrow n}^{rel},
	\label{RhoDiagMaster}
\end{align} 
where the transition rates $\Gamma_{n\rightarrow k}$ are given by Eq.~(\ref{GammaRel}).

{\it Lindblad form.} 
Eqs.~(\ref{RhoOffDiagMaster}) and (\ref{RhoDiagMaster}) for the evolution
of the density matrix can be rewritten in the Lindblad form
\begin{align}
	\partial_t \hrho=-i[\cH_{\text{eff}},\hrho]
	-\frac{1}{2}\sum_{i,j}
	\left(
	\cL_{ij}^\dagger\cL_{ij}\hrho
	+\hrho\cL_{ij}^\dagger\cL_{ij}
	-2\cL_{ij}^\dagger\hrho\cL_{ij}
	\right),
	\label{Lindblad}
\end{align}
where the summation runs over all pairs of indices $i={1,\ldots,N}$ and $j={1,\ldots,N}$ in an $N$-level system;
the effective Hamiltonian of coherent evolution is given by
\begin{align}
	\cH_{\text{eff}}=\sum_i \ket{i}\bra{i}(E_i+\delta E_i),
\end{align}
and the Lindblad operators
\begin{align}
	\cL_{ij}=(1-\delta_{ij})\sqrt{\Gamma_{j\rightarrow i}^{rel}}\ket{i}\bra{j}
	+\delta_{ij}\sqrt{S(0)/N}\sum_l V_{ll}\ket{l}\bra{l}
\end{align}
account for the effects of dephasing and dissipation.

\subsection{Master equations for OTOCs}

In what follows we derive microscopically the Bloch-Redfield-type
master equations for the out-of-time-order correlator (\ref{Kdef}), following a procedure similar to
the derivation (see, e.g., Ref.~\cite{Slichter:book}) of the master equations for the density matrix. 
Due to the weakness of the system-environment coupling,
the OTOCs decay on long times significantly exceeding the characteristic correlation time
of the environment.

It follows directly from Eq.~(\ref{Kdef}) that
\begin{align}
	\partial_t K_{m_1m_2,n_1n_2}= 
	i\left<\left[\cH_0+\cH_{\text{coupl}}(t),\ket{n_1}\bra{m_1}(t)\right]\hB(0)\ket{n_2}\bra{m_2}(t)\hD(0)\right>
	\nonumber\\
	+i\left<\ket{n_1}\bra{m_1}(t)\hB(0)\left[\cH_0+\cH_{\text{coupl}}(t),\ket{n_2}\bra{m_2}(t)\right]\hD(0)\right>,
	\label{preMaster}
\end{align}
where 
$\cH_0$ is the Hamiltonian of the system (without the environment) and
$\cH_{\text{coupl}}=\hX\sum_{n,m}V_{nm}\ket{n}\bra{m}$ is the coupling between the system and the environment.
By expanding all Heisenberg operators in Eq.~(\ref{preMaster}) to
the first order in the perturbation $\cH_{\text{coupl}}$
and neglecting the change of the density matrix of the system during the characteristic 
correlation time of the environment, we arrive at the equations for the evolution of the elements  
$K_{m_1m_2,n_1n_2}$ in the form
\begin{align}
	\partial_t K_{m_1m_2,n_1n_2}= 
	&i(E_{n_1}+E_{n_2}-E_{m_1}-E_{m_2})K_{m_1m_2,n_1n_2}
	\nonumber\\
	&-\left<\int_{-\infty}^t\left[\cH_{\text{coupl}}(t^\prime),\left[\cH_{\text{coupl}}(t),\ket{n_1}\bra{m_1}(t)\right]\right]dt^\prime\,\,\hB(0)\ket{n_2}\bra{m_2}(t)\,\,\hD(0)\right>
	\nonumber\\
	&-\left<\ket{n_1}\bra{m_1}(t)\,\hB(0)\int_{-\infty}^t\left[\cH_{\text{coupl}}(t^\prime),
	\left[\cH_{\text{coupl}}(t),\ket{n_2}\bra{m_2}(t)\right]\right]dt^\prime\,\,\hD(0)\right>
	\nonumber\\
	&-\left<\left[\cH_{\text{coupl}}(t),\ket{n_1}\bra{m_1}(t)\right]\hB(0)
	\int_{-\infty}^t\left[\cH_{\text{coupl}}(t^\prime),\ket{n_2}\bra{m_2}(t)\right]dt^\prime\,\,\hD(0)\right>
	\nonumber\\
	&-\left<\int_{-\infty}^t\left[\cH_{\text{coupl}}(t^\prime),\ket{n_1}\bra{m_1}(t)\right]dt^\prime\,\,\hB(0)
	\left[\cH_{\text{coupl}}(t),\ket{n_2}\bra{m_2}(t)\right]\hD(0)\right>,
	\label{MasterGeneric}
\end{align}
where only the terms up to the second order in the system-environment coupling have been kept
and the lower time integration limit has been extended to $-\infty$ in view of the short correlation
time of the environment degrees of freedom, i.e.
the correlation time between $\cH_{\text{coupl}}(t^\prime)\propto \hX(t^\prime)$
and $\cH_{\text{coupl}}(t)\propto \hX(t)$.
Using Eq.~(\ref{MasterGeneric}), we derive below the master
equations for the evolution of the OTOCs in the form (\ref{KelemEvol}).

Due to the weakness of the system-environment coupling,
the characteristic energy gaps between system levels significantly exceed the decay rates of the
OTOCs, which are determined by the last four lines in Eq.~(\ref{MasterGeneric}); the elements
$K_{m_1m_2,n_1n_2}$ quickly oscillate with frequencies $E_{n_1}+E_{n_2}-E_{m_1}-E_{m_2}$
and decay with rates significantly exceeded by these frequencies.
Thus, the evolution of each element $K_{m_1m_2,n_1n_2}$ depends only on other elements corresponding to the
same energy splitting $E_{n_1}+E_{n_2}-E_{m_1}-E_{m_2}$. Below we consider separately the cases of finite
and zero values of the splitting.

\subsubsection{Finite energy splitting}

For each combination of different $m_1$, $m_2$, $n_1$ and $n_2$ there are four elements $K$ which correspond
to the same energy splitting and differ from each other by permutations of indices. We assume for simplicity that
there is no additional degeneracy of the quantities $E_{n_1}+E_{n_2}-E_{m_1}-E_{m_2}$ when all of the indices
$m_1$, $m_2$, $n_1$ and $n_2$ are different.
Eq.~(\ref{MasterGeneric}) in that case gives
\begin{align}
	\partial_t K_{m_1m_2,n_1n_2}=
	& i(E_{n_1}+\delta E_{n_1}+E_{n_2}+\delta E_{n_2}
	-E_{m_1}-\delta E_{m_1}-E_{m_2}-\delta E_{m_2})K_{m_1m_2,n_1n_2}
	\nonumber \\
	& -\frac{1}{2}\left(\sum_{k\neq m_1}\Gamma_{m_1\rightarrow k}^{rel}
	+\sum_{k\neq n_1}\Gamma_{n_1\rightarrow k}^{rel}
	+\sum_{k\neq m_2}\Gamma_{m_2\rightarrow k}^{rel}
	+\sum_{k\neq n_2}\Gamma_{n_2\rightarrow k}^{rel}
	\right)K_{m_1m_2,n_1n_2}
	\nonumber \\
	&-\Gamma_{n_2\rightarrow n_1}^{rel} K_{m_1m_2,n_2n_1}
	-\Gamma_{m_1\rightarrow m_2}^{rel} K_{m_2m_1,n_1n_2}
	-\Gamma^{\phi}_{m_1n_1,m_2n_2} K_{m_1m_2,n_1n_2},
	\label{KelemMasterDifferent}
\end{align}
where the transition rates $\Gamma_{i\rightarrow j}^{rel}$ are given by Eq.~(\ref{GammaRel});
$\delta E_i$ is the renormalisation of the $i$-th level by environment, given by Eq.~(\ref{LambShift}); and
\begin{align}
	\Gamma^{\phi}_{m_1n_1,m_2n_2}=\frac{1}{2}(V_{n_1n_1}+V_{n_2n_2}-V_{m_1m_1}-V_{m_2m_2})^2S(0)
	\label{DephCollective}
\end{align}
is the dephasing rate in a compound system consisting of two copies of the original system coupled to
the same bath.

%\tr{(Warning: when some of indices conincide, but the splitting is still finite, additional
%terms arise due to additional degeneracies, like for elements $K_{km_2,n_1k}$. We do not consider this case
%here.)}

\subsubsection{Zero energy splitting}

Elements $K_{m_1m_2,n_1n_2}$ with zero splitting $E_{n_1}+E_{n_2}-E_{m_1}-E_{m_2}$ have a greater degeneracy and
require separate analyses.

{\it ``Diagonal'' elements.}
Let us first consider the elements with $n_1=m_1$ and $n_2=m_2$.
These elements satisfy the same equations of evolution as the diagonal elements
of the density matrix of a compound system consisting of two copies of the original system.
For $n_1=n_2=n\neq m=m_1=m_2$ we obtain from Eq.~(\ref{MasterGeneric})
\begin{align}
	\partial_t K_{nm,nm}=
	& -K_{nm,nm}\sum_{k\neq n} \Gamma_{n\rightarrow k}^{rel}
	-K_{nm,nm}\sum_{k\neq m} \Gamma_{m\rightarrow k}^{rel}
	%\nonumber\\
	+\sum_{k\neq n} \Gamma_{k\rightarrow n}^{rel}K_{km,km}
	+\sum_{k\neq m} \Gamma_{k\rightarrow m}^{rel}K_{nk,nk}
	\nonumber\\
	&-\Gamma_{m\rightarrow n}^{rel}K_{nm,mn}
	-\Gamma_{n\rightarrow m}^{rel}K_{mn,nm}.
	\label{DiagEvol1}
\end{align}
In the case $n=m$ Eq.~(\ref{MasterGeneric}) gives
\begin{align}
	\partial_t K_{nn,nn}=&-2K_{nn,nn}\sum_{k\neq n} \Gamma_{n\rightarrow k}^{rel}
	%\nonumber\\	&
	+\sum_{k\neq n} \left(\Gamma_{k\rightarrow n}^{rel}K_{kn,kn}
	+\Gamma_{k\rightarrow n}^{rel}K_{nk,nk}\right)
	\nonumber\\
	&+\sum_{k\neq n} \left(\Gamma_{k\rightarrow n}^{rel}K_{nk,kn}
	+\Gamma_{n\rightarrow k}^{rel}K_{kn,nk}\right).
	\label{DiagEvol2}
\end{align}
From Eqs.~(\ref{DiagEvol1}) and (\ref{DiagEvol2}) it follows immediately
that
\begin{align}
	\sum_{m,n} K_{mm,nn}=\text{const},
	\label{DiagConservSM}
\end{align}
which corresponds to the conservation of the sum of the diagonal elements of the density matrix of a compound system.

{\it ``Non-diagonal'' elements.} The other set of elements with zero energy splitting,
different from the ``diagonal'' elements, correspond to $m_1=n_2$ and $m_2=n_1$. Their evolution is described
by the equations
\begin{align}
	\partial_t K_{mn,nm}=
	%\nonumber\\
	&-\left(\sum_{k\neq m}\Gamma_{m\rightarrow k}^{rel}+\sum_{k\neq n}\Gamma_{n\rightarrow k}^{rel}\right)K_{mn,nm}
	%\nonumber\\	&
	-\left(K_{mn,mn}+K_{nm,nm}\right)\Gamma_{m\rightarrow n}^{rel}
	\nonumber\\
	&+\sum_{k\neq m}K_{kn,nk}\Gamma_{m\rightarrow k}^{rel}
	+\sum_{k\neq n}K_{mk,km}\Gamma_{k\rightarrow n}^{rel}.
\end{align}

\subsection{Master equation for the density matrix for two copies of a system coupled to the same environment}

The equations for the evolution of the elements $K_{m_1m_2,n_1n_2}$ are similar to the equations of evolution
of the density-matrix elements $\rho_{m_1m_2,n_1n_2}=\left<\ket{n_1n_2}\bra{m_1m_2}(t)\right>$
of a compound system consisting of two copies
of the original system coupled to the same environment, where $n_i$ and $m_i$
label the states of the $i$-th subsystem; $i=1,2$.
The Hamiltonian of such a compound system is given by Eq.~(\ref{HamTwoSyst}).
%\begin{align}
%	\cH=\cH_0\otimes\holOne + \holOne\otimes\cH_0
%	+\left(\hV\otimes\holOne + \holOne\otimes\hV\right)\hX+\cH_{\text{bath}}(\hat X),
%\end{align}
To the second order in the system-environment coupling $\hV$
the evolution of the density matrix elements is described by the equation
\begin{align}
	\partial_t \rho_{m_1m_2,n_1n_2}= 
	&i(E_{n_1}+E_{n_2}-E_{m_1}-E_{m_2})\rho_{m_1m_2,n_1n_2}
	\nonumber\\
	&-\left<\int_{-\infty}^t\left[\hX(t^\prime)\hV(t^\prime)\otimes\holOne + \holOne\otimes\hV(t^\prime)\hX(t^\prime),
	\left[\hX(t)\hV(t)\otimes\holOne + \holOne\otimes\hV(t)\hX(t),
	\ket{n_1n_2}\bra{m_1m_2}(t)\right]\right]dt^\prime\right>,
	\label{RhoTwoGenEvol}
\end{align}
The form of the coupling $\hV=\sum_{n,m}V_{nm}\ket{n}\bra{m}$ and Eq.~(\ref{RhoTwoGenEvol})
give, when all of the indices $n_1$, $n_2$, $m_1$ and $m_2$ are different,
\begin{align}
	\partial_t \rho_{m_1m_2,n_1n_2}=
	& i(E_{n_1}+\delta E_{n_1}+E_{n_2}+\delta E_{n_2}
	-E_{m_1}-\delta E_{m_1}-E_{m_2}-\delta E_{m_2})\rho_{m_1m_2,n_1n_2}
	\nonumber \\
	& -\frac{1}{2}\left(\sum_{k\neq m_1}\Gamma_{m_1\rightarrow k}^{rel}
	+\sum_{k\neq n_1}\Gamma_{n_1\rightarrow k}^{rel}
	+\sum_{k\neq m_2}\Gamma_{m_2\rightarrow k}^{rel}
	+\sum_{k\neq n_2}\Gamma_{n_2\rightarrow k}^{rel}
	\right)\rho_{m_1m_2,n_1n_2}
	\nonumber \\
	&-\frac{1}{2}\left(\Gamma_{n_2\rightarrow n_1}^{rel}+\Gamma_{n_1\rightarrow n_2}^{rel}
	+iE_{n_1n_2}^{flip}\right)\rho_{m_1m_2,n_2n_1}
	\nonumber \\
	&-\frac{1}{2}\left(\Gamma_{m_2\rightarrow m_1}^{rel}+\Gamma_{m_1\rightarrow m_2}^{rel}
	-iE_{m_1m_2}^{flip}\right)\rho_{m_2m_1,n_1n_2}
	\nonumber\\
	&-\Gamma^{\phi}_{m_1n_1,m_2n_2} \rho_{m_1m_2,n_1n_2},
	\label{RhoMasterTwoDifferent}
\end{align}
where the quantity
\begin{align}
	E_{n_1n_2}^{flip}=|V_{n_1n_2}|^2\int\frac{d\omega}{2\pi}
	\frac{S(\omega)-S(-\omega)}{\omega+E_{n_1n_2}}
	\label{FlipFlopRate}
\end{align}
gives the rate of the flip-flop processes, i.e. the rate of the coherent interchange $n_1 \leftrightarrow n_2$,
and the dephasing rate $\Gamma^{\phi}_{m_1n_1,m_2n_2}$ is defined by Eq.~(\ref{DephCollective}).

{\it Lindblad form.} The master equations for the evolution of the density matrix of two systems in the same environment
may may be also rewritten in the Lindblad form (\ref{Lindblad}) with the effective Hamiltonian
\begin{align}
	\cH_{\text{eff}}=\sum_i \ket{i}\bra{i}
	\left[
	(E_i+\delta E_i)\otimes\holOne
	+\holOne\otimes(E_i+\delta E_i)
	\right]
	+\frac{1}{2}\sum_{i,j}E_{ij}^{flip}\ket{i}\bra{j}\otimes\ket{j}\bra{i}
\end{align}
and the Lindblad operators
\begin{align}
	\cL_{ij}=(1-\delta_{ij})\sqrt{\Gamma_{j\rightarrow i}^{rel}}
	\left(\ket{i}\bra{j}\otimes\holOne+\holOne\otimes\ket{i}\bra{j}\right)
	+\delta_{ij}\sqrt{S(0)/N}\sum_l V_{ll}\left(\ket{l}\bra{l}\otimes\holOne+\holOne\otimes\ket{l}\bra{l}\right).
\end{align}

{\it Mapping between OTOCs and two-system density matrix.} Eq.~(\ref{KelemMasterDifferent}), which described
the evolution of OTOCs for an open system in a dissipative environment, resembles Eq.~(\ref{RhoMasterTwoDifferent}),
which describes the evolution of the density matrix elements for two copies of the system coupled to this
environment. Indeed, both equations have the same diagonal part, i.e. the part which relates the evolution
of the element $\rho_{m_1m_2,n_1n_2}$ or $K_{m_1m_2,n_1n_2}$ to itself. Both equations also 
have terms with interchanged indices $n_1\leftrightarrow n_2$ or $m_1\leftrightarrow m_2$.
While two systems coupled to an environment allow for a coherent (``flip-flop'') as well as inelastic
interchange, the respective processes for OTOCs
are purely inelastic.

As discussed in the main text, in the limit of a classical environment the evolutions
of the OTOC and two systems coupled to this environment may be mapped onto each other. 
Classical environment corresponds to the odd spectrum $S(\omega)=S(-\omega)$, which leads to
the vanishing of the flip-flop rates (\ref{FlipFlopRate}) and identical relaxation rates 
$\Gamma_{n_1\rightarrow n_2}^{rel}=\Gamma_{n_2\rightarrow n_1}^{rel}$ of the transitions 
$n_1\rightarrow n_2$ and $n_2\rightarrow n_1$ for each pair of states $n_1$ and $n_2$.
The equations (\ref{KelemMasterDifferent}) and (\ref{RhoMasterTwoDifferent}) for the evolution of the OTOC
and the two systems become identical in this limit.

\end{document}